\documentclass[aps,prl,twocolumn,showpacs,groupedaddress,amsmath,amssymb]{revtex4}

\usepackage[dvips]{graphicx}% Include figure files
\usepackage{dcolumn}% Align table columns on decimal point
\usepackage{bm}% bold math
\begin{document}

\preprint{APS/123-QED}

\title{Mode Selection in the Spontaneous Motion of an Alcohol Droplet}

\author{Ken Nagai}
 \affiliation{Department of Physics, Graduate School of Science, Kyoto University, Kyoto 606-8502, Japan}

\author{Yutaka Sumino}
 \affiliation{Department of Physics, Graduate School of Science, Kyoto University, Kyoto 606-8502, Japan}

\author{Hiroyuki Kitahata}
 \affiliation{Department of Physics, Graduate School of Science, Kyoto University, Kyoto 606-8502, Japan}

\author{Kenichi Yoshikawa}
 \email[To whom correspondence should be addressed. Tel:+81-75-753-3812. Fax:+81-75-753-3779. Email:]{yoshikaw@scphys.kyoto-u.ac.jp}
 \affiliation{Department of Physics, Graduate School of Science, Kyoto University, Kyoto 606-8502, Japan}

\date{\today}

\begin{abstract}
An alcohol (pentanol) droplet exhibits spontaneous agitation on an aqueous solution, driven by a solutal Marangoni effect. We found that the droplet's mode of motion is controlled by its volume. A droplet with a volume of less than $0.1\; \mu\rm{l}$ shows irregular translational motion, whereas intermediate-sized droplets of $0.1-200\; \mu\rm{l}$ show vectorial motion. When the volume is above $300\; \mu\rm{l}$, the droplet splits into smaller drops. These experimental results regarding mode selection are interpreted in terms of the wave number selection depending on the droplet volume.  
\end{abstract}

% PACS, the Physics and Astronomy
                             % Classification Scheme.
%\keywords{Suggested keywords}%Use showkeys class option if keyword
                              %display desired
\maketitle

It is well known that an oil-water system exhibits spontaneous agitation, called the Marangoni effect \cite{ME}, which is driven either by a thermal gradient \cite{TM} or a chemical concentration gradient \cite{CM1,CM2,CM3}. We focus here on the latter system, i.e., a solutal Marangoni effect under isothermal conditions. A rich variety of spontaneous agitation on oil-water interface has been reported both experimentally \cite{OW2,OW3,Cho,CMEnergy-trans,PFOA,Free-Run,BZ,NE,NE2} and theoretically \cite{OWT1,OWT2}. It has been shown that the nature of the self-agitation is sensitively dependent on the shape and size of the container, i.e., on the boundary condition \cite{OW2,OW3,Cho}.  In the present study, we show that an alcohol droplet driven by a solutal Marangoni effect selects a certain mode of motion critically depending on its size. This mode selection is interpreted in terms of competition between the droplet size and the critical wave number in the instability by the Marangoni effect. 

\begin{figure}
\includegraphics{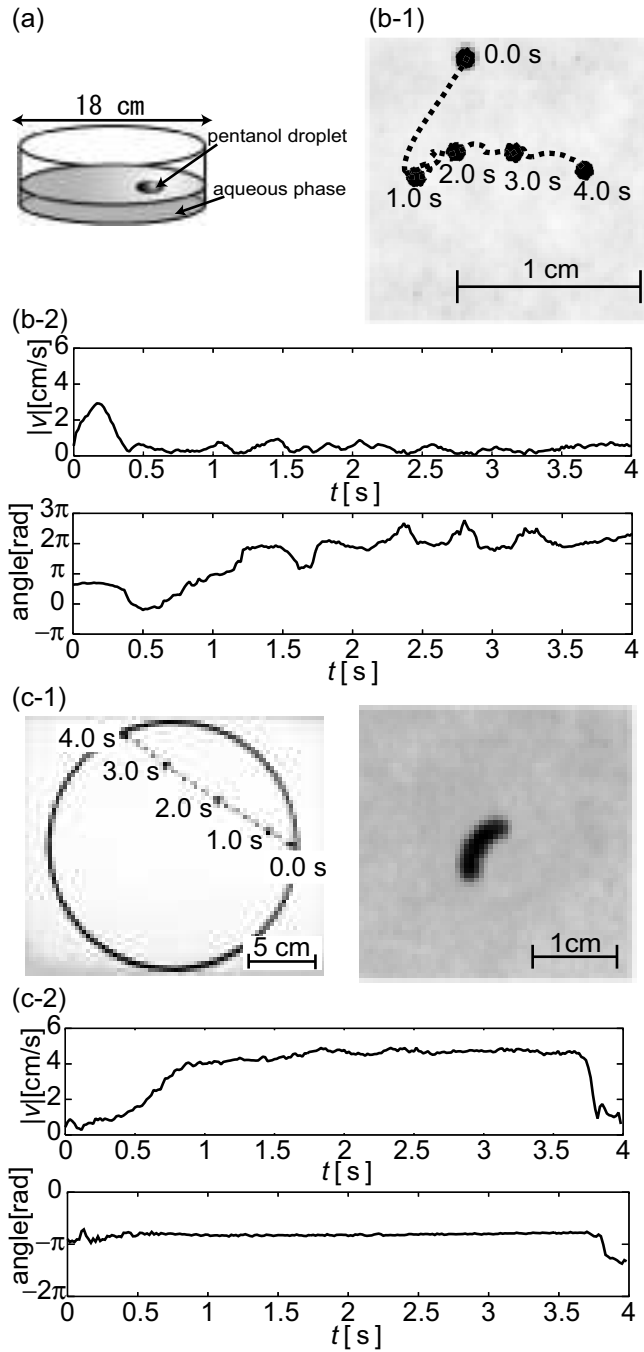}
\caption{(a) Schematic illustration of the experimental setup. (b-1) Irregular translational motion of a droplet with a volume of $0.017\; \mu\rm{l}$. Images of the droplet at every 1 second are shown on the broken line, which shows the trace of droplet motion. (b-2) Time trace of the speed and direction of motion. (c-1) (left) Vectorial motion of a droplet with a volume of $10\; \mu\rm{l}$. Images of the droplet at every 1 second are shown on the broken line, which shows the trace of droplet motion. (right) Morphology of the droplet at $2.0\; \rm{s}$. (c-2) Time trace of the speed and direction of motion.}
\label{small}
\end{figure}

\begin{figure}
\includegraphics{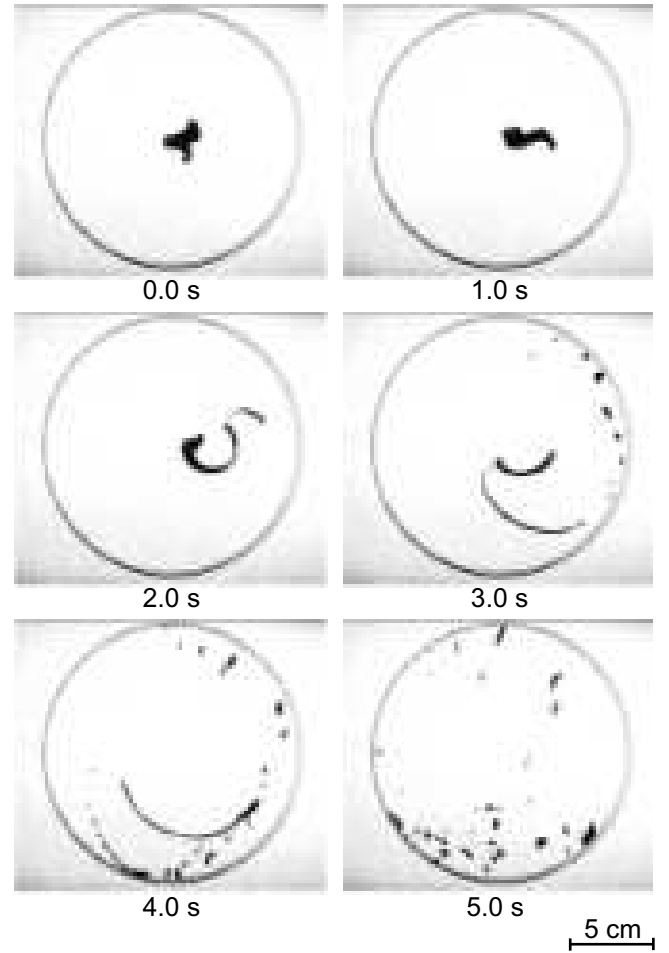}
\caption{Spontaneous motion of a droplet with a volume of $400\; \mu\rm{l}$, exhibiting fission into small droplets.}
\label{ookii}
\end{figure}

\begin{figure}
\includegraphics{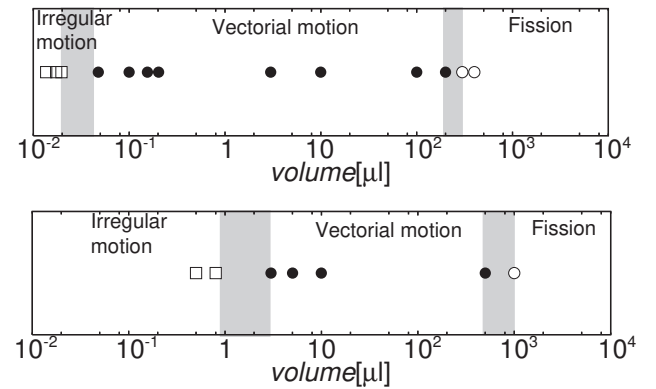}
\caption{Phase diagram of the mode of droplet motion. (upper) The alcohol concentration of the aqueous phase is $2.3\; \rm{vol}\%$. (lower) The alcohol concentration of the aqueous phase is $2.4\; \rm{vol}\%$. Irregular translational motion at $\Box$. Vectorial motion at $\bullet$. Fission into smaller droplets at $\circ$. When the volume was less than $0.1\; \mu\rm{l}$, we hypothesize that the volume is proportional to the $3/2$ power of the cross-section of the droplet. In the shaded regions, the mode of the droplet motion is not decisive to a single specific mode.}
\label{phase}
\end{figure}

Different modes of spontaneous motion of alcohol droplets are exemplified in Figs. \ref{small} and \ref{ookii}, where the droplets stained with ink (Pilot Corporation, Tokyo; INKSP-55-B) are agitating on $100\; \rm{ml}$ of aqueous phase containing $2.3\; \rm{vol}\%$ pentanol in a petri dish with a diameter of $18\; \rm{cm}$. The motion of the droplet was monitored with a high-speed video camera (RedLake MASD Inc., San Diego; Motion Scope PCI) at 60 frames per second at room temperature, and then analyzed by image-processing software. When the volume is less than $0.1\; \mu\rm{l}$, the droplet maintains a circular shape during irregular translational motion, as shown in Fig. \ref{small}(b). When the volume is between 0.1 and $200\; \mu\rm{l}$, the droplet shows an asymmetric morphology and exhibits directional motion by maintaining its shape, as in Fig. \ref{small}(c), where the speed is almost constant. The average speed in vectorial motion is greater than that in the irregular translational motion of the smaller droplet. When the volume is above $300\; \mu\rm{l}$, the droplet splits into smaller droplets as in Fig. \ref{ookii}. Figure \ref{phase} shows a phase diagram of the modes of droplet motion with a change in the volume. 


\begin{figure}
\includegraphics{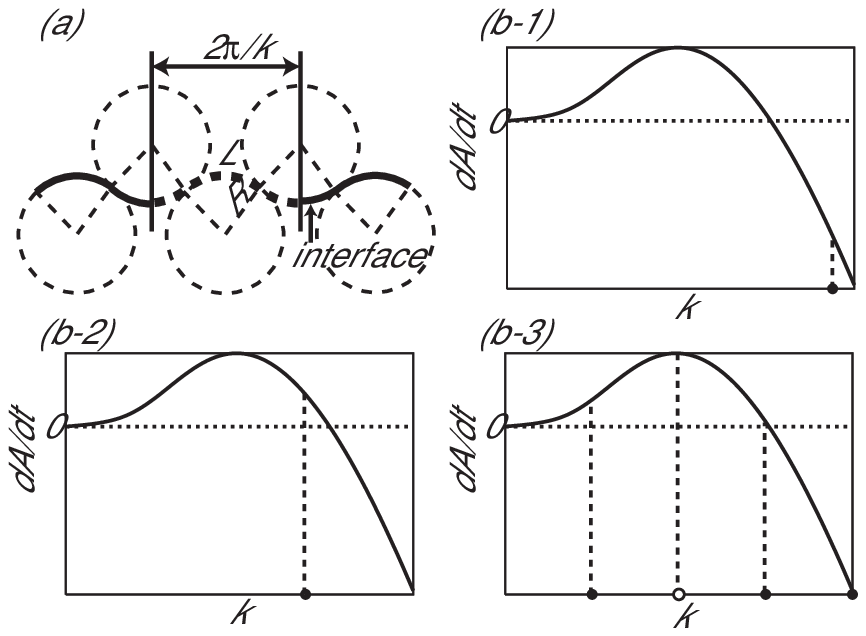}
\caption{(a) Schematic illustration of the tractable model. (b) Relationship between the wave number of Marangoni instability and the size of a droplet. Closed and open circles ($\bullet$,$\circ$) correspond to the eigenvalues of the sine wave around the droplet. (b-1) When there are no eigenvalues where the wave grows, the droplet cannot break symmetry and show irregular translational motion due to the small fluctuation along the periphery. (b-2) When only the fundamental wave is slightly unstable, the droplet exhibits vectorial motion by keeping an asymmetric morphology, as in Fig. \ref{small}(c-1). (b-3) When the wave number of the maximum instability ($\circ$) grows on a large enough droplet, fission into smaller droplets occurs.}
\label{en}
\end{figure}

The experimental trend in volume-dependent mode-selection can be interpreted by considering the wave-length due to the Marangoni instability and the length scale of the droplet. When a droplet is large enough to accept the characteristic length of Marangoni instability, it exhibits morphological deformation. Once the droplet is deformed, the variation in curvature around its periphery causes asymmetry in the surface tension, which drives the directional motion of the droplet. In the experiment in Fig. \ref{small}(c), the concentration gradient of pentanol in the convex region of the droplet is higher than that in the concave region of the droplet due to diffusion depending on the shape of the boundary. Therefore, the interfacial tension gradient in the convex region is higher than that in the concave region and the droplet is forced to move toward the direction from the concave region to the convex region \cite{Cho}. This spontaneous motion has the effect of maintaining the shape anisotropy. In contrast, in the case of a smaller droplet, the circular shape is maintained under Marangoni instability, and the emergence of a large acceleration force is inhibited. As a result, the average speed of such a circular droplet is less than that of a larger droplet and the spontaneous agitation is rather irregular. The interface is significantly destabilized only when the wave number of the self-agitation due to the Marangoni effect is less than the inverse of a certain wave number, $1/k_{\rm{c}}$. This mode selection is shown schematically in Fig. \ref{en}(b).

Next, we calculate the wave number when the interface is destabilized by a solutal Marangoni effect by considering a small perturbation ($y=A\sin kx$, $|A|\ll 1$). When a wave is induced along the interface, generally the wave tends to be damped due to interfacial tension, the strength of which per unit length is $\gamma_{1}$. On the other hand, the concentration gradient in the convex region of the wave is higher than that in the concave region due to diffusion depending on the shape of the boundary. Therefore, the interfacial tension gradient in the convex region is higher than that in the concave region and the difference in the interfacial tension gradient induces wave growth. The force strength per unit length is $\gamma_{2}$. When $\gamma_{2}>\gamma_{1}$, the wave becomes unstable. To simplify the treatment, we adopt the approximation that the shape of the wave consists of parts of a circle, as in Fig. \ref{en}(a). 

In this case, the radius of the circle, $R$, and the length of the interface per wave length, $L$, are derived as 
\begin{equation}
R = \frac{A^2 + \left( {\pi / 2k} \right)^2}{2A} \approx \frac{\pi ^2}{8Ak^2},
\end{equation}
\begin{equation}
L = 4R\arcsin \left( {\frac{\pi / 2k}{R}} \right) \approx  2\frac{\pi 
}{k}\left( {1 + \frac{8A^2k^2}{3\pi ^2}} \right).
\end{equation}
Supposing that $\gamma_{1}$ is proportional to $\partial \left[L/\left( 2\pi /k\right)\right]/\partial A$,
\begin{equation}
\gamma_{1}=a\frac{\partial}{\partial A}\left(\frac{L}{2\pi/k}\right)=aAk^{2},
\end{equation}
where $a$ is a constant.

\begin{figure}
\includegraphics{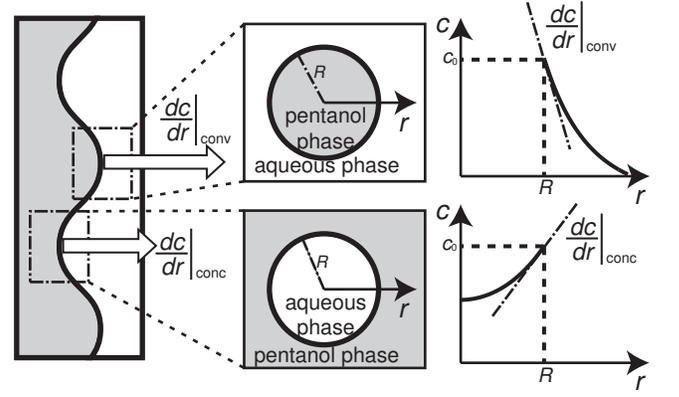}
\caption{Schematic representation on the interfacial tension difference. As the interface is assumed to be circular with the radius of $R$, the concentration profile is $\left( c_{0}-\Gamma \right)K_{0}( \sqrt{\left( \alpha + \beta \right)/D}r  )/K_{0}( \tilde{R})$ at the convex region (upper), and $\left( c_{0}-\Gamma \right)I_{0}( \sqrt{\left( \alpha + \beta \right)/D}r )/I_{0}( \tilde{R})$ at the concave region (lower). The interfacial tension is proportional to the gradient of the concentration at the interface; see text.}
\label{dcdr}
\end{figure}

$\gamma_{2}$ is proportional to the difference between the interfacial tension gradients in the convex and concave regions. Assuming that the interfacial tension gradients are proportional to the concentration gradients of pentanol,
\begin{equation}
\gamma_{2}=b\left[ \left| \left( \left. \frac{dc}{dr}\right|_{\rm{conv}} \right) \right| -\left| \left( \left. \frac{dc}{dr}\right|_{\rm{conc}} \right) \right| \right],
\end{equation}
where $\left. dc/dr\right|_{\rm{conv}}$ is the concentration gradient perpendicular to the interface in the convex region, $\left. dc/dr\right|_{\rm{conc}}$ is that in the concave region, and $b$ is a constant. The dynamics of the concentration of pentanol on the surface of the aqueous phase, $c$, are given by
\begin{equation}
\frac{\partial c}{\partial t}=D\nabla^{2}c-\alpha \left( c-c_{\rm{a}}\right) -\beta \left( c-c_{\rm{w}}\right),
\label{kakusan}
\end{equation}
where $D$ is the diffusion coefficient of pentanol on the surface, $\alpha$ is the evaporation rate of pentanol, $\beta$ is the dissolution rate of pentanol, $c_{\rm{a}}$ is the concentration at the gas-liquid equilibrium and $c_{\rm{w}}$ is the concentration at the surface-bulk equilibrium in the aqueous phase. We impose the Dirichlet condition at the pentanol-water interface, i.e., $c=c_{0}$ at the interface, where $c_{0}$ is the concentration of the water-rich part under the coexistence condition. The stationary solutions with rotational symmetry of Eq. \ref{kakusan} (Fig. \ref{dcdr}) are adopted in order to evaluate $\gamma_{2}$. Thus, at the convex region,
\begin{equation}
\left| \left( \left .\frac{dc}{dr}\right|_{\rm{conv}} \right) \right| = \left( c_0-\Gamma\right) \sqrt{\frac{\alpha+\beta}{D}} \frac{K_1 \left( \tilde{R} \right)}{K_0 \left( \tilde{R}\right)},
\end{equation}
and at the concave region,
\begin{equation}
\left| \left( \left. \frac{dc}{dr}\right|_{\rm{conc}} \right) \right| = \left( c_0-\Gamma\right) \sqrt{\frac{\alpha+\beta}{D}} \frac{I_1 \left(\tilde{R} \right)}{I_0 \left( \tilde{R} \right)},
\end{equation}
where $\Gamma=\left( \alpha c_{\rm{a}}+\beta c_{\rm{w}}\right)/\left( \alpha +\beta\right)$, $\tilde{R}=\sqrt{\left( \alpha+\beta\right)/D}R$ and $I_{0}$, $I_{1}$, $K_{0}$, $K_{1}$ are modified Bessel functions. Therefore,
\begin{eqnarray}
\gamma_{2}&=&b\left[ \left| \left( \left. \frac{dc}{dr}\right|_{\mathrm{conv}} \right) \right| - \left| \left( \left. \frac{dc}{dr}\right|_{\mathrm{conc}} \right) \right| \right]\nonumber\\
&=&  b\left( c_0-\Gamma \right) \sqrt{\frac{\alpha+\beta}{D}} \left( \frac{K_1 \left( \tilde{R} \right)}{K_0 \left( \tilde{R} \right)}-\frac{I_1 \left( \tilde{R} \right)}{I_0 \left( \tilde{R} \right)} \right).
\end{eqnarray}

Here, we assume $0<kA=\delta \ll 1$. When $k\ll 1$,
\begin{equation}
\gamma_{2}-\gamma_{1} \approx  -aAk^{2}+\frac{8b\Gamma Ak^{2}}{\pi ^{2}}, 
\end{equation}
and when $k\gg 1$,
\begin{eqnarray}
\gamma_{2}-\gamma_{1} &\approx& -aAk^{2} \nonumber\\
	&&+\frac{8b\Gamma Ak^{2}}{\pi ^{2}\ln \left[ 8Ak^{2}/\left( \pi ^{2} \sqrt{\left( \alpha+\beta \right)/D}\right) \right] }.
\end{eqnarray}
This means $\gamma_{2}-\gamma_{1}<0$ when $k\gg 1$, and $\gamma_{2}-\gamma_{1}>0$ when $k\ll 1$. Since $\gamma_{2}-\gamma_{1}=0$ at $k=k_{\rm{c}}$, 
\begin{equation}
k_{\rm{c}}=\frac{\pi^{2}}{8\delta}\sqrt{\frac{\alpha+\beta}{D}}\exp \left[\frac{3b\left( c_{0}-\Gamma \right)}{2a}\right].
\end{equation}

From the above discussion, it is clear that the perturbation increases when $k<k_{\rm{c}}$. This means that for larger $\Gamma$, $k_{\rm{c}}$ is smaller, i.e., the critical volume at the mode transition increases. 

To verify this analysis, we performed experiments where $\Gamma$ was changed by covering the dish and increasing the pentanol concentration in the aqueous phase. When the dish was covered and $\Gamma$ increased, the motion of a droplet with a volume of $1.3\; \mu\rm{l}$ changed from vectorial motion to irregular translational motion (data not shown). When the concentration of the aqueous phase was $2.4\; \rm{vol}\%$ and $\Gamma$ increased, the critical volume at which the mode change is induced increases as shown in Fig. \ref{phase}. These results correspond well to our theoretical expectation.

In the present study, we found that mode selection in the spontaneous motion of an alcohol droplet is induced by a change in the droplet volume. This mode selection is achieved as the result of the competition in the length scale between the droplet size and the wavelength of the instability due to the Marangoni effect. The characteristic wavelength is a function of rates of evaporation and dissolution. The present results may shed light on how biological motors realize chemo-mechanical energy transduction under isothermal conditions with high efficiency \cite{Bio,Bio1,Bio2}.

 We would like to thank Mr. M. I. Kohira (Chuo University, Japan) and Prof. M. Nagayama (Kanazawa University, Japan) for their helpful discussion and advice. This work was supported in part by a Grant-in-Aid for the 21st Century COE (Center for Diversity and Universality in Physics) from the Ministry of Education, Culture, Sports, Science and Technology of Japan.


\begin{thebibliography}{21}
\expandafter\ifx\csname natexlab\endcsname\relax\def\natexlab#1{#1}\fi
\expandafter\ifx\csname bibnamefont\endcsname\relax
  \def\bibnamefont#1{#1}\fi
\expandafter\ifx\csname bibfnamefont\endcsname\relax
  \def\bibfnamefont#1{#1}\fi
\expandafter\ifx\csname citenamefont\endcsname\relax
  \def\citenamefont#1{#1}\fi
\expandafter\ifx\csname url\endcsname\relax
  \def\url#1{\texttt{#1}}\fi
\expandafter\ifx\csname urlprefix\endcsname\relax\def\urlprefix{URL }\fi
\providecommand{\bibinfo}[2]{#2}
\providecommand{\eprint}[2][]{\url{#2}}


\bibitem[{\citenamefont{Scriven and Sternling}(1960)}]{ME}
\bibinfo{author}{\bibfnamefont{L.~E.} \bibnamefont{Scriven}} \bibnamefont{and}
  \bibinfo{author}{\bibfnamefont{C.~V.} \bibnamefont{Sternling}},
  \bibinfo{journal}{Nature} \textbf{\bibinfo{volume}{187}},
  \bibinfo{pages}{186} (\bibinfo{year}{1960}).

\bibitem[{\citenamefont{Schwabe et~al.}(1992)\citenamefont{Schwabe,
  M{\"{o}}ller, Schneider, and Scharmann}}]{TM}
\bibinfo{author}{\bibfnamefont{D.}~\bibnamefont{Schwabe}},
  \bibinfo{author}{\bibfnamefont{U.}~\bibnamefont{M{\"{o}}ller}},
  \bibinfo{author}{\bibfnamefont{J.}~\bibnamefont{Schneider}},
  \bibnamefont{and}
  \bibinfo{author}{\bibfnamefont{A.}~\bibnamefont{Scharmann}},
  \bibinfo{journal}{Phys. Fluids A} \textbf{\bibinfo{volume}{4}},
  \bibinfo{pages}{2368} (\bibinfo{year}{1992}).

\bibitem[{\citenamefont{Kovalchuk et~al.}(1999)\citenamefont{Kovalchuk,
  Kamusewitz, Vollhardt, and Kovalchuk}}]{CM1}
\bibinfo{author}{\bibfnamefont{V.~I.} \bibnamefont{Kovalchuk}},
  \bibinfo{author}{\bibfnamefont{H.}~\bibnamefont{Kamusewitz}},
  \bibinfo{author}{\bibfnamefont{D.}~\bibnamefont{Vollhardt}},
  \bibnamefont{and} \bibinfo{author}{\bibfnamefont{N.~M.}
  \bibnamefont{Kovalchuk}}, \bibinfo{journal}{Phys. Rev. E}
  \textbf{\bibinfo{volume}{60}}, \bibinfo{pages}{2029} (\bibinfo{year}{1999}).

\bibitem[{\citenamefont{Kovalchuk and Vollhardt}(2000)}]{CM2}
\bibinfo{author}{\bibfnamefont{N.~M.} \bibnamefont{Kovalchuk}}
  \bibnamefont{and}
  \bibinfo{author}{\bibfnamefont{D.}~\bibnamefont{Vollhardt}},
  \bibinfo{journal}{J. Phys. Chem. B} \textbf{\bibinfo{volume}{104}},
  \bibinfo{pages}{7987} (\bibinfo{year}{2000}).

\bibitem[{\citenamefont{Kovalchuk and Vollhardt}(2004)}]{CM3}
\bibinfo{author}{\bibfnamefont{N.~M.} \bibnamefont{Kovalchuk}}
  \bibnamefont{and}
  \bibinfo{author}{\bibfnamefont{D.}~\bibnamefont{Vollhardt}},
  \bibinfo{journal}{Phys. Rev. E} \textbf{\bibinfo{volume}{69}},
  \bibinfo{pages}{016307} (\bibinfo{year}{2004}).

\bibitem[{\citenamefont{Yoshikawa and Magome}(1993)}]{OW2}
\bibinfo{author}{\bibfnamefont{K.}~\bibnamefont{Yoshikawa}} \bibnamefont{and}
  \bibinfo{author}{\bibfnamefont{N.}~\bibnamefont{Magome}},
  \bibinfo{journal}{Bull. Chem. Soc. Jpn.} \textbf{\bibinfo{volume}{66}},
  \bibinfo{pages}{3352} (\bibinfo{year}{1993}).

\bibitem[{\citenamefont{Sumino et~al.}(2005)\citenamefont{Sumino, Magome,
  Hamada, and Yoshikawa}}]{OW3}
\bibinfo{author}{\bibfnamefont{Y.}~\bibnamefont{Sumino}},
  \bibinfo{author}{\bibfnamefont{N.}~\bibnamefont{Magome}},
  \bibinfo{author}{\bibfnamefont{T.}~\bibnamefont{Hamada}}, \bibnamefont{and}
  \bibinfo{author}{\bibfnamefont{K.}~\bibnamefont{Yoshikawa}},
  \bibinfo{journal}{Phys. Rev. Lett.} \textbf{\bibinfo{volume}{94}},
  \bibinfo{pages}{068301} (\bibinfo{year}{2005}).

\bibitem[{\citenamefont{Nakata et~al.}(1997)\citenamefont{Nakata, Iguchi, Ose,
  Kuboyama, Ishii, and Yoshikawa}}]{Cho}
\bibinfo{author}{\bibfnamefont{S.}~\bibnamefont{Nakata}},
  \bibinfo{author}{\bibfnamefont{Y.}~\bibnamefont{Iguchi}},
  \bibinfo{author}{\bibfnamefont{S.}~\bibnamefont{Ose}},
  \bibinfo{author}{\bibfnamefont{M.}~\bibnamefont{Kuboyama}},
  \bibinfo{author}{\bibfnamefont{T.}~\bibnamefont{Ishii}}, \bibnamefont{and}
  \bibinfo{author}{\bibfnamefont{K.}~\bibnamefont{Yoshikawa}},
  \bibinfo{journal}{Langmuir} \textbf{\bibinfo{volume}{13}},
  \bibinfo{pages}{4454} (\bibinfo{year}{1997}).

\bibitem[{\citenamefont{Kitahata and Yoshikawa}(in press)}]{CMEnergy-trans}
\bibinfo{author}{\bibfnamefont{H.}~\bibnamefont{Kitahata}} \bibnamefont{and}
  \bibinfo{author}{\bibfnamefont{K.}~\bibnamefont{Yoshikawa}},
  \bibinfo{journal}{Physica D}  (\bibinfo{year}{in press}).

\bibitem[{\citenamefont{Bain et~al.}(1994)\citenamefont{Bain, Burnett-Hall, and
  Montgomerie}}]{PFOA}
\bibinfo{author}{\bibfnamefont{C.~D.} \bibnamefont{Bain}},
  \bibinfo{author}{\bibfnamefont{G.~D.} \bibnamefont{Burnett-Hall}},
  \bibnamefont{and} \bibinfo{author}{\bibfnamefont{R.~R.}
  \bibnamefont{Montgomerie}}, \bibinfo{journal}{Nature}
  \textbf{\bibinfo{volume}{372}}, \bibinfo{pages}{414} (\bibinfo{year}{1994}).

\bibitem[{\citenamefont{Santos et~al.}(1995)\citenamefont{Santos, and 
  Ondar\c{c}uhu}}]{Free-Run}
\bibinfo{author}{\bibfnamefont{F.~D.~D.} \bibnamefont{Santos}},
  \bibnamefont{and} \bibinfo{author}{\bibfnamefont{T.} \bibnamefont{Ondar\c{c}uhu}},
  \bibinfo{journal}{Phys. Rev. Lett.}
  \textbf{\bibinfo{volume}{75}}, \bibinfo{pages}{2972} (\bibinfo{year}{1995}).

\bibitem[{\citenamefont{Kitahata et~al.}(2002)\citenamefont{Kitahata, Aihara,
  Magome, and Yoshikawa}}]{BZ}
\bibinfo{author}{\bibfnamefont{H.}~\bibnamefont{Kitahata}},
  \bibinfo{author}{\bibfnamefont{R.}~\bibnamefont{Aihara}},
  \bibinfo{author}{\bibfnamefont{N.}~\bibnamefont{Magome}}, \bibnamefont{and}
  \bibinfo{author}{\bibfnamefont{K.}~\bibnamefont{Yoshikawa}},
  \bibinfo{journal}{J. Chem. Phys.} \textbf{\bibinfo{volume}{116}},
  \bibinfo{pages}{5666} (\bibinfo{year}{2002}).

\bibitem[{\citenamefont{Bekki et~al.}(1990)\citenamefont{Bekki, Vignes-Adler,
  Nakache, and Adler}}]{NE}
\bibinfo{author}{\bibfnamefont{S.}~\bibnamefont{Bekki}},
  \bibinfo{author}{\bibfnamefont{M.}~\bibnamefont{Vignes-Adler}},
  \bibinfo{author}{\bibfnamefont{E.}~\bibnamefont{Nakache}}, \bibnamefont{and}
  \bibinfo{author}{\bibfnamefont{P.~M.} \bibnamefont{Adler}},
  \bibinfo{journal}{J. Colloid Interface Sci.} \textbf{\bibinfo{volume}{140}},
  \bibinfo{pages}{492} (\bibinfo{year}{1990}).

\bibitem[{\citenamefont{Bekki et~al.}(1992)\citenamefont{Bekki, Vignes-Adler,
  and Nakache}}]{NE2}
\bibinfo{author}{\bibfnamefont{S.}~\bibnamefont{Bekki}},
  \bibinfo{author}{\bibfnamefont{M.}~\bibnamefont{Vignes-Adler}},
  \bibnamefont{and} \bibinfo{author}{\bibfnamefont{E.}~\bibnamefont{Nakache}},
  \bibinfo{journal}{J. Colloid Interface Sci.} \textbf{\bibinfo{volume}{152}},
  \bibinfo{pages}{314} (\bibinfo{year}{1992}).

\bibitem[{\citenamefont{de~Genne}(1998)}]{OWT1}
\bibinfo{author}{\bibfnamefont{P.~G.} \bibnamefont{de~Genne}},
  \bibinfo{journal}{Physica A} \textbf{\bibinfo{volume}{249}},
  \bibinfo{pages}{196} (\bibinfo{year}{1998}).

\bibitem[{\citenamefont{Thiele et~al.}(2004)\citenamefont{Thiele, John, and
  B\"{a}r}}]{OWT2}
\bibinfo{author}{\bibfnamefont{U.}~\bibnamefont{Thiele}},
  \bibinfo{author}{\bibfnamefont{K.}~\bibnamefont{John}}, \bibnamefont{and}
  \bibinfo{author}{\bibfnamefont{M.}~\bibnamefont{B\"{a}r}},
  \bibinfo{journal}{Phys. Rev. Lett.} \textbf{\bibinfo{volume}{93}},
  \bibinfo{pages}{027802} (\bibinfo{year}{2004}).

\bibitem[{\citenamefont{Shr{\"{o}}dinger}(1944)}]{Bio}
\bibinfo{author}{\bibfnamefont{E.}~\bibnamefont{Shr{\"{o}}dinger}},
  \emph{\bibinfo{title}{What is Life?}} (\bibinfo{publisher}{Cambridge
  University Press}, \bibinfo{address}{Cambridge}, \bibinfo{year}{1944}).

\bibitem[{\citenamefont{Mehta et~al.}(1999)\citenamefont{Mehta, Rock, Rief,
  Spudich, Mooseker, and Cheney}}]{Bio1}
\bibinfo{author}{\bibfnamefont{A.~D.} \bibnamefont{Mehta}},
  \bibinfo{author}{\bibfnamefont{R.~S.} \bibnamefont{Rock}},
  \bibinfo{author}{\bibfnamefont{M.}~\bibnamefont{Rief}},
  \bibinfo{author}{\bibfnamefont{J.~A.} \bibnamefont{Spudich}},
  \bibinfo{author}{\bibfnamefont{M.~S.} \bibnamefont{Mooseker}},
  \bibnamefont{and} \bibinfo{author}{\bibfnamefont{R.~E.}
  \bibnamefont{Cheney}}, \bibinfo{journal}{Nature}
  \textbf{\bibinfo{volume}{400}}, \bibinfo{pages}{590} (\bibinfo{year}{1999}).

\bibitem[{\citenamefont{Yasuda et~al.}(2001)\citenamefont{Yasuda, Noji,
  Yoshida, {Kinoshita Jr.}, and Itoh}}]{Bio2}
\bibinfo{author}{\bibfnamefont{R.}~\bibnamefont{Yasuda}},
  \bibinfo{author}{\bibfnamefont{H.}~\bibnamefont{Noji}},
  \bibinfo{author}{\bibfnamefont{M.}~\bibnamefont{Yoshida}},
  \bibinfo{author}{\bibfnamefont{K.}~\bibnamefont{{Kinoshita Jr.}}},
  \bibnamefont{and} \bibinfo{author}{\bibfnamefont{H.}~\bibnamefont{Itoh}},
  \bibinfo{journal}{Nature} \textbf{\bibinfo{volume}{410}},
  \bibinfo{pages}{898} (\bibinfo{year}{2001}).
\end{thebibliography}
\end{document}